\documentclass[12pt,preprint]{aastex}
\usepackage{epsf}

\begin{document} 

\title{Detecting the Dusty Debris of Terrestrial Planet Formation}

\author{Scott J. Kenyon}
\affil{Smithsonian Astrophysical Observatory, 60 Garden Street, 
Cambridge, MA 02138, USA; e-mail: skenyon@cfa.harvard.edu}
\author{and}
\author{Benjamin C. Bromley}
\affil{Department of Physics, University of Utah, 201 JFB, Salt Lake City, 
UT 84112, USA; e-mail: bromley@physics.utah.edu}

\begin{abstract}
We use a multiannulus accretion code to investigate debris disks 
in the terrestrial zone, at 0.7--1.3 AU around a 1 $M_{\odot}$ star. 
Terrestrial planet formation produces a bright dusty ring of debris
with a lifetime of $\gtrsim 10^6$ yr.  The early phases of terrestrial 
planet formation are observable with current facilities; the late 
stages require more advanced instruments with adaptive optics.
\end{abstract}

\subjectheadings{planetary systems -- solar system: formation -- 
stars: formation -- circumstellar matter}

\section{INTRODUCTION}

Stars are born with circumstellar disks.  Young stars have opaque, 
gaseous disks with radii of 100--1000 AU and masses of 
0.001--0.1 $M_{\odot}$ \citep[e.g.][]{wya03}.  
The disks absorb and re-emit light from the central 
star and generate internal energy by viscous dissipation
\citep[e.g.,][]{lbp74,ada87,kh87,ber88,chi99}.  
Typical disk temperatures range from $\sim$ 2000~K at the 
inner edge to $\sim$ 30--50~K at the outer edge.
Because the disk is cooler than the star, it produces 
an infrared (IR) excess of radiation compared to a normal
stellar photosphere.
The excess is $\sim$ 1 mag at 3.5 $\mu$m (L-band), 2--5 mag at
10 $\mu$m (N-band), and 8--10 mag at 100 $\mu$m
\citep[e.g.,][]{kh95}.

As they age, stars lose their disks.  The fraction of
stars with opaque disks declines from 80\%--90\% for 
1 Myr old stars to 10\%--20\% for 10 Myr old stars
\citep{hai01b}.
Stars older than 10 Myr rarely have opaque disks.
However, many older stars have optically thin, `debris disks,'
with radii of 100--1000 AU and masses of 0.01--10 $M_{\oplus}$
\citep[e.g.,][]{hab01,spa01,gre03}.
Debris disks are cooler than opaque disks and emit most of 
their radiation at wavelengths, $\lambda \gtrsim$ 10 $\mu$m
\citep[e.g.,][]{lag00}.

Recent observations suggest a rapid evolution from opaque 
disk to debris disk. In the Taurus-Auriga cloud, many young
stars have the mid-IR colors of an opaque disk, K--L $\sim$ 1, 
K--N $\gtrsim$ 2; others have the colors of cool main--sequence 
stars, K--L $\lesssim$ 0.4, K--N $\lesssim$ 0.5 \citep{kh95,har90}. 
Only $\sim$ 1\%--2\% of the 1--3 Myr old stars in this cloud 
have intermediate colors. Other nearby star-forming regions 
have similar fractions of `transition' disks
\citep[e.g.][]{bon01,hai01a,jay01,sta01,naj03},
suggesting a transition timescale of $\lesssim 10^5$ yr
for material at 0.1--3 AU.  

The transformation from opaque disk to debris disk requires 
the `disappearance' of $\sim$ 0.001--0.1 $M_{\odot}$ of gas and 
dust. Planet formation plausibly explains this evolution.  
In the planetesimal model, dust grains grow to mm sizes and fall 
into the disk midplane.  The grains evolve into 1~km `planetesimals' 
\citep{gol73,wei93,you03}, which collide and merge into planets
\citep[e.g.][]{lis93,ws93,kl99}.
The gravity of 1000~km or larger planets stirs up leftover 
planetesimals to large velocities; a cascade of collisions 
then removes most of the solid material remaining in small 
objects \citep{wil94}. At the same time, accretion, photoevaporation, 
disk and stellar winds, and perhaps other processes remove 
the gas \citep[e.g.][]{hol94,pol96,kon00}.

Recent studies support this picture.  
Observations suggest grain growth in the disks surrounding
the youngest stars \citep[e.g.][]{DA01,thr01,hog03,vB03}. 
Vortices and other density fluctuations in the disk promote the 
growth of grains into planetesimals \citep{ada95,sup00,hag03}.
At $\sim$ 0.3--3 AU, rocky planetesimals merge and grow 
into planets in 1--10 Myr \citep{ws93,wei97,cha01}.
At 10--50 AU, icy planetesimals become planets in 10--30 Myr 
\citep[][]{kl99}.  Once icy planets form, a collisional
cascade explains the luminosities, structures, and timescales
of debris disks \citep[][2004]{dom04,ken99,kb02b}.

Here, we consider the formation of terrestrial planets and
debris disks.  A collisional cascade associated with the
formation of 2000~km and larger rocky planets produces measurable 
IR excesses which last for $\sim 10^6$ yr, comparable to transition 
times observed in pre-main--sequence stars.  The early stages 
of terrestrial planet formation are observable with current
facilities. The final accumulation phases require more sensitive 
instruments with adaptive optics.

We outline the model in \S2, describe the calculations in \S3,
and conclude with a brief discussion in \S4.

\section{THE MODEL}

We adopt the \citet{saf69} statistical approach to calculate the 
collisional evolution of an ensemble of planetesimals in orbit
around a star of mass $M_{\odot}$ and luminosity $L_{\odot}$
\citep[][2004]{kb02a}.  The model grid contains $N$ concentric 
annuli with widths $\delta a_i$ centered at heliocentric distances 
$a_i$.  Calculations begin with a differential mass distribution 
$n(m_{ik}$) of bodies with orbital eccentricity $e_{ik}(t)$ and 
inclination $i_{ik}(t)$.

To evolve the mass and velocity distributions in time, we solve 
the coagulation and Fokker-Planck equations for bodies undergoing 
inelastic collisions, drag forces, and long-range gravitational 
interactions \citep{kb02a}.  We adopt collision rates from kinetic
theory and use an energy-scaling algorithm to assign collision 
outcomes \citep{ws93,wei97,dav85}.  
To compute collision rates and outcomes between planetesimals in
different annuli, we derive an overlap region whose size depends
on the eccentricities and masses of the planetesimals in each 
annulus \citep[][Appendix A]{kb02a}.
We derive changes in orbital parameters from gas drag, 
dynamical friction, and viscous stirring \citep{ada76, oht02}.  
Our approach does not include tidal interactions between the largest
bodies and the gas \citep[e.g.][]{agn02} or the largest bodies and
smaller planetesimals \citep[][2003]{raf01}. Our tests show that 
these effects are small for the early evolution discussed here, 
but become important at later stages when the largest bodies 
merge into planets. We plan to discuss these issues in future 
papers.  \citet[][2002a]{kb01} summarize algorithms and tests 
of the code \citep[see also][1999]{kl98}.

To compute dust masses and luminosities, we use dust production
rates and particle scale heights derived from the multiannulus 
coagulation code as input to a separate dust evolution code \citep{kb04}. 
This second calculation includes collisions among dust grains along 
with gas and Poynting-Robertson drag.  
Because collision timescales are shorter than timescales for gas and 
radiation drag, collisions set the size distribution for small grains 
and the total optical depth through the disk.
We use a simple radiative transfer method to derive the optical
depth through the disk; dust luminosities follow from the dust scale
height.  To derive dust temperatures, we assume grains have an albedo 
of 20\% and emit as blackbodies in radiative equilibrium.

\section{CALCULATIONS}

Our numerical calculations begin with 1--1000 m planetesimals in
32 concentric annuli centered at a distance of 1 AU from the Sun. 
We consider (i) thin torus calculations at 0.84-1.16 AU with 
$\delta a_i$ = 0.01 and (ii) thick torus calculations at 0.68-1.32 AU 
with $\delta a_i$ = 0.02.  
The planetesimals have an initial surface density 
$\Sigma = \Sigma_0$ (a/1~AU)$^{-3/2}$.  We consider models with
$\Sigma_0$ = 8, 16, and 32 g cm$^{-2}$. The `minimum mass solar nebula'
has $\Sigma_0 \sim$ 16 g cm$^{-2}$ \citep{wei77,hay81}.  In each
annulus, the initial mass distribution has 30--60 mass batches, with
a mass spacing, $\delta = m_{i+1}/m_i$ = 1.4--2.0, between successive 
batches.  The cumulative number of bodies is $N_C \propto m_i^{-0.03}$.  
We add mass batches as planetesimals grow in mass.  
Each planetesimal batch begins in a nearly circular orbit with 
eccentricity $e_0 = 10^{-4}$ and inclination $i_0 = e_0/2$.  The 
initial gas density at 1 AU is $\rho_g$ = $1.07 \times 10^{-9}$ 
($\Sigma_0$/16 g cm$^{-2}$) g cm$^{-3}$; \citet{kb02a} describe how 
we scale the gas density with radius and time.  For fragmentation
parameters, we adopt a tensile strength in the range 
$S_0 = 10^6$--$10^8$ erg g$^{-1}$ and a crushing energy 
$Q_c = 10^7$--$10^9$ erg g$^{-1}$ \citep{dav85,ws93}. 

In our calculations, the growth of planetesimals into planets follows 
a standard pattern \citep{lis93,ws93,wei97}. At 1 AU,
it takes 500--1000 yr for 1~km objects to grow into 100~km objects.
Collisions produce a small amount of debris, which adds material 
to the mass batches for 1--100~m objects. Gas drag circularizes 
the orbits of small objects and removes $\sim$ 0.001\% of this 
mass from the innermost annuli.  Viscous stirring raises the
eccentricities of the 0.1--10~km objects; dynamical friction 
lowers the eccentricities of larger objects.  Gravitational
focusing factors increase and runaway growth begins.

Runaway growth concentrates most of the mass in the largest objects.
It takes $\sim 1-2 \times 10^4$ yr to produce 1000~km objects 
and another $\sim 1-2 \times 10^4$ yr to produce 2000~km objects.  
During runaway growth, viscous stirring raises the velocities of 100~m 
to 10~km bodies to the disruption limit. Collisions among these
bodies then produce substantial debris instead of mergers. Continued
stirring lowers gravitational focusing factors and retards the
growth of the largest objects.  As large objects grow slowly, 
erosive collisions reduce the number of 100~m to 10~km objects 
substantially.

Figure 1 shows two snapshots of the mass distribution for large 
objects in a thin torus calculation. During the first $10^5$ yr,
protoplanets form in the inner annuli. As the collisional cascade 
begins at $\sim$ 0.9 AU, protoplanets start to form at $\sim$ 1.1 AU.  
Smaller gravitational focusing factors and depletion by erosive 
collisions then limit growth.  By $\sim 10^6$ yr, planet growth 
saturates at radii $\sim$ 3000~km.  A few protoplanets each with 
masses of 0.01--0.1 $M_{\earth}$ then contain $\gtrsim$ 70\% of the 
remaining mass (25\%--50\% of the initial mass).  The large eccentricities 
of these protoplanets allow their orbits to cross; multiple 
orbit-crossings lead to mergers and the formation of Earth-like 
planets \citep{cha01}. Because our 
coagulation code does not treat these interactions accurately, we 
stopped our calculations at $\sim$ 2 Myr. We plan to report on 
calculations that incorporate an $n$-body code in future papers 
\citep{bk04}.

Figure 2 shows the evolution of the dust mass for models with
\citet{ws93} fragmentation. We divide the debris into small grains 
with sizes of 1--1000 $\mu$m and large particles with sizes of 
1--1000 mm. At the start of these calculations, 
collisions yield mergers with little dust. As larger objects 
form, their gravity stirs up planetesimals to the disruption 
velocity. Dust production grows rapidly. The dust mass peaks when 
the largest objects have radii of $\sim$ 2000~km. Because the 
growth time varies inversely with the initial mass in 
planetesimals, more massive disks reach peak dust mass before
less massive disks. More massive disks also produce more dust. 
We derive a simple relation for the dust mass as 
a function of initial surface density and time:
\begin{equation}
M_{S/L} \approx M_{S/L,0} ~ \left ( \frac{\Sigma_0}{\rm 16 ~ g ~ cm^{-2}} \right ) \left ( \frac{t_0}{t} \right ) ~~~~~ {\rm for} ~~ t \gtrsim t_0,
\end{equation}
where $M_{S,0} \sim 10^{23}$ g for small particles, 
$M_{L,0} \sim 3 \times 10^{24}$ g for large particles,
and $t_0 \sim 10^5$ yr.
This relation is fairly independent of the fragmentation 
algorithm and the material properties of the planetesimals
\citep{kb01,dom04}.

Despite the rapid decline in dust mass, dust in the terrestrial 
zone produces an observable IR excess.  Figure 3 illustrates the 
evolution of the 10 $\mu$m and 20 $\mu$m excesses relative to a 
normal stellar photosphere for models with \citet{ws93} fragmentation.
The IR excesses peak when the dust mass reaches its maximum. The 
largest planets then have radii of 2000~km.  As planets grow 
to sizes of 3000~km and larger, IR excesses decline considerably.  
The 0.01--0.1 $M_{\oplus}$ objects then collide and merge into 
Earth-like planets \cite[][]{cha01,bk04}.  During this final 
accumulation phase, 10--20 $\mu$m excesses are small.   

As the 10--20 $\mu$m excesses decline, individual collisions 
among 10--100 km objects produce fluctuations in the dust 
production rate. These disruptive collisions yield large
variations in the 10--20 $\mu$m excesses.

Infrared excesses are relatively insensitive to the size of the 
zone where terrestrial planets form.  In the left panels of Figure 3, 
calculations in a thin torus at 0.84-1.16 AU produce excesses 
of $\sim$ 1--2 mag at 10 $\mu$m and $\sim$ 3 mag at 20 $\mu$m. 
Calculations in a larger torus yield somewhat larger excesses 
(Figure 3; right panels). We estimate that a complete calculation 
of the terrestrial zone at 0.3--3 AU would yield 10--20 $\mu$m 
excesses $\lesssim$ 1 mag larger than the excesses in Figure 3.

Our calculations indicate that the magnitude of the 10--20 $\mu$m 
excess produced by terrestrial dust is insensitive to many model 
parameters.  Because collisions dominate the destruction of 
planetesimals and dust grains, the details of gas and 
Poynting-Robertson drag are unimportant.  Rapid early growth 
erases the initial size distribution; variations in the slope 
of the initial power-law size distribution do not modify 
the results.  Changes in the fragmentation parameters $S_0$ and 
$Q_c$ produce modest, $\sim$ 25\%, changes in the dust production 
rates and the growth timescales. As noted in previous studies, 
\citet{dav85} fragmentation yields shorter growth times and smaller 
dust production rates; this algorithm also concentrates a larger
fraction of the initial mass into the largest bodies. However, 
the long-term evolution of the IR excesses in models with 
\citet{dav85} fragmentation is nearly identical to the evolution 
with \citet{ws93} fragmentation \citep{kb04}. 

The initial mass in planetesimals controls the timescale of 
the dust evolution.  Massive disks produce larger planets more 
rapidly than low mass disks.
The timescale for the formation of 2000--3000~km planets scales 
inversely with the mass \citep[][1999]{lis87,ws93,wei97,kl98}, 
$t_P \approx 1-2 \times 10^5$~yr 
$( {\rm 16 ~ g ~ cm^{-2} / \Sigma_0} )$. 

The initial surface density also sets the geometry of the dusty disk.  
In calculations with $\Sigma_0 \gtrsim$ 16 g cm$^{-2}$, the
radial optical depth of the disk often exceeds unity. The opaque, 
inner disk can then shadow the outer disk \citep{kb04}. 
Shadowing reduces the disk luminosity and the magnitude of the
IR excess. Thus, massive disks are sometimes less luminous
than low mass disks (Figure 3).

Comparisons with other approaches support our estimates.
\citet{woo02} assume the disk mass decreases homologously with 
time and derive IR excesses from detailed radiative transfer 
calculations. Their predictions for 10 $\mu$m excesses, 
$\lesssim$ 2 mag for disks with $\lesssim$ $10^{24}$ g in
1--100 $\mu$m particles at 0.2--2 AU, agree with our results.

\section{SUMMARY}

Our calculations yield the first demonstration that terrestrial
planet formation produces observable amounts of dust. The 10 $\mu$m 
excesses range from $\sim$ 2 mag at $t \sim 10^5$ yr to 
$\lesssim$ 0.25 mag at $t \sim 10^6$ yr. The excesses are
largest when 1000--2000~km objects form; IR excesses fade as 
protoplanets collide and begin to merge into Earth-mass planets.

The size and duration of IR excesses from terrestrial planet
formation are similar to values needed to explain objects in
transition from the opaque disks of pre-main--sequence stars
to optically thin debris disks.  Our models yield maximum IR 
excesses of N--N$_0$ $\approx$ 2 and Q--Q$_0$ $\approx$ 3--4; 
T Tauri stars with disks have N--N$_0$ $\approx$ 2--5 and Q--Q$_0$ 
$\approx$ 4--8.  In $\sim 10^6$ yr, the model IR excesses 
decline to values consistent with those of T Tauri stars 
without opaque disks, N--N$_0$ $\lesssim$ 0.4 and 
Q--Q$_0$ $\lesssim$ 0.5. IR excesses probably remain at 
this level for $\sim 1-10 \times 10^7$ yr, as terrestrial 
planet formation concludes.

Quantifying IR excesses from 300 K dust around 1--100 Myr old 
solar-type stars tests predictions for dust masses and timescales 
for the formation of 1000--3000~km protoplanets.  The models 
predict observable 10--20 $\mu$m excesses for $\lesssim$ 1 Myr 
old stars and small excesses for older stars. Current ground-based 
10--20 $\mu$m data agree with this prediction \citep[e.g.,][]{wei03}.  
Data from {\it SIRTF} should provide better statistics on IR 
excesses and a better test of the model.
During the final accumulation phase, IR excesses are too small for 
detection with direct imaging observations.  However, instruments 
with adaptive optics on 10--100~m class telescopes can resolve 
the radial dust distribution and provide direct constraints on 
the late epochs of terrestrial planet formation.
For terrestrial planets forming around 1--10 Myr old stars, we
expect dusty structures on 0.1--1 AU scales \citep[see also][]{oze00}, 
roughly similar to the 10--100 AU structures observed in debris 
disks around nearby A-type stars \citep[e.g.][]{jay98,sch99,wei99,wil02}.
Direct detections of these structures around nearby solar-type stars
would provide good evidence for recent formation of terrestrial planets.

\vskip 6ex

We acknowledge a generous allotment, $\sim$ 1250 cpu days, of 
computer time at the supercomputing center at the Jet Propulsion 
Laboratory through funding from the NASA Offices of Mission to 
Planet Earth, Aeronautics, and Space Science.  
Advice and comments from M. Geller, A. Weinberger, and the 
referee improved our presentation.  
The {\it NASA} {\it Astrophysics Theory Program} supported 
part of this project through grant NAG5-13278.

\clearpage


\centerline{Figure Captions}

\noindent
{Fig. 1 -- Masses of the largest objects for a thin torus model with
$\Sigma_0$ = 16 g cm$^{-2}$ and \citet{ws93} fragmentation.
(a) lower panel: $t = 10^5$ yr;
(b) upper panel: $t = 10^6$ yr.
The horizontal error bars indicate the extent of 
the orbit for each object.}

\noindent
{Fig. 2 -- Evolution of the dust mass in 
small grains ('S'; 1 $\mu$m $\le r \le$ 1 m) and
large grains ('L'; 1 mm $\le r \le$ 1 m) for models with
$\Sigma_0$ as indicated in the legend. More massive disks 
produce more debris at earlier times than less massive 
disks. For all disks, the dust production rate converges 
to a power-law decline with time.}

\noindent
{Fig. 3 -- Evolution of the broadband 10 $\mu$m (N-N$_0$) and 
20 $\mu$m (Q-Q$_0$) excesses as a function of time. 
Left panels: 0.84--1.16 AU torus calculations;
right panels: 0.68--1.32 AU torus calculations.
Fragmentation produces large excesses relative to
a stellar photosphere at $t \sim 10^4$ to $10^5$ yr.
By $t \sim 10^6$ yr, the 10 $\mu$m excess is nearly 
unobservable; a small excess persists at 20 $\mu$m.}
\clearpage

\begin{figure}
\plotone{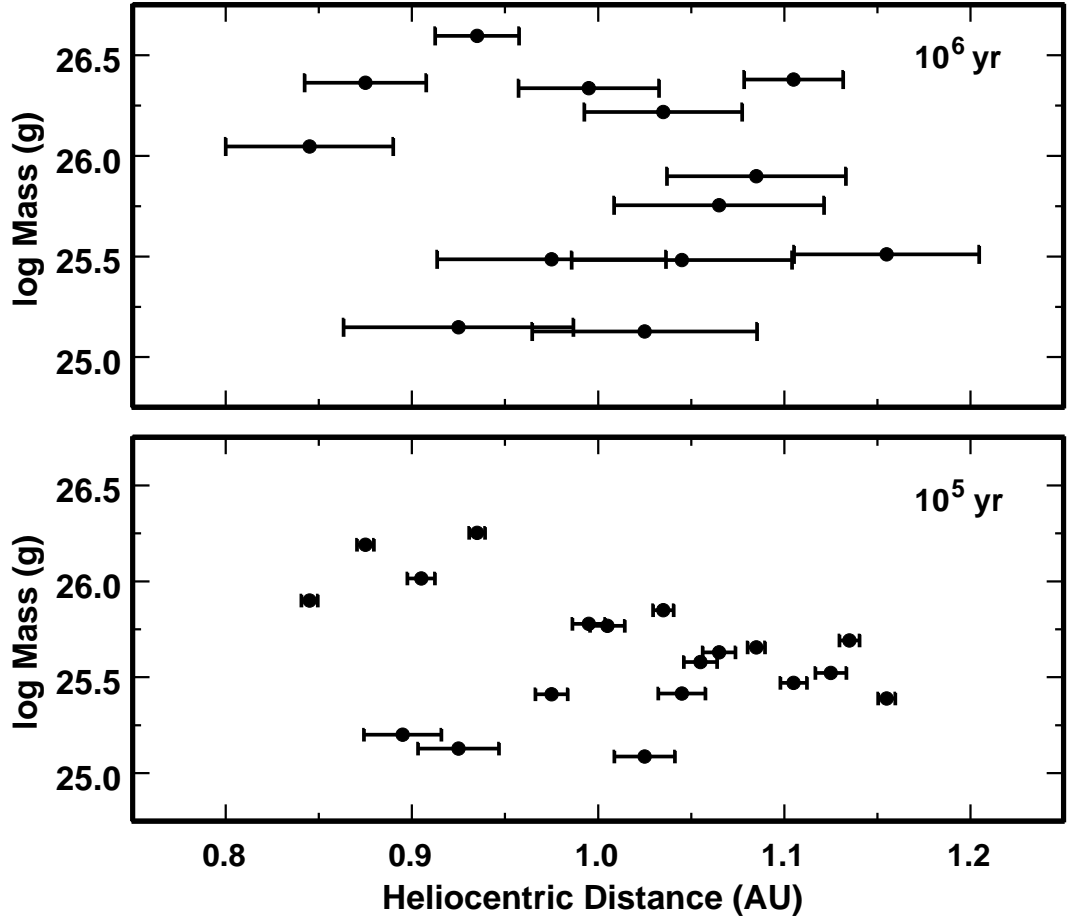}
\caption
{Masses of the largest objects for a thin torus model with
$\Sigma_0$ = 16 g cm$^{-2}$ and \citet{ws93} fragmentation.
(a) lower panel: $t = 10^5$ yr;
(b) upper panel: $t = 10^6$ yr.
The horizontal error bars indicate the extent of
the orbit for each object.}
\end{figure}
\clearpage

\begin{figure}
\plotone{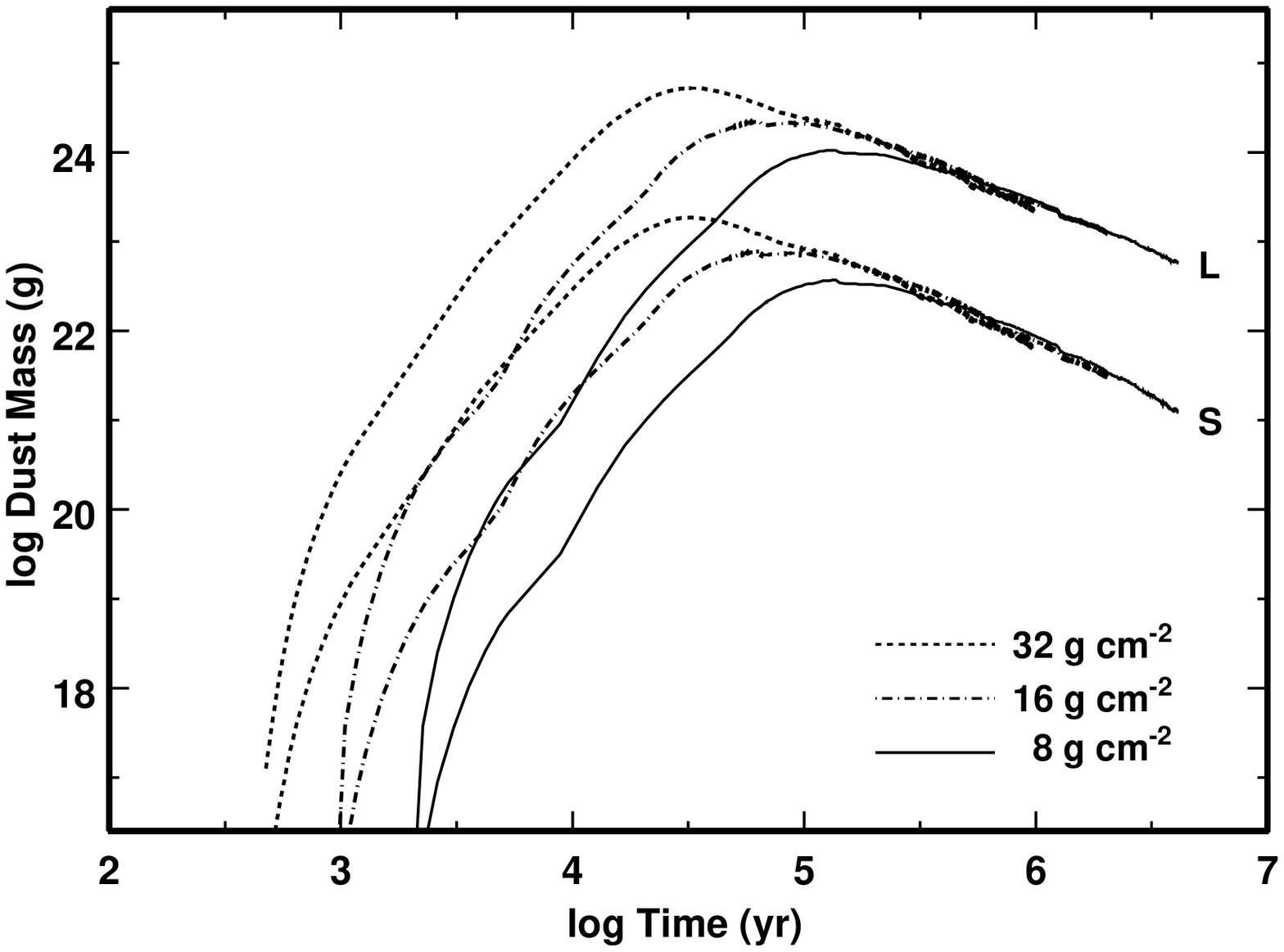}
\caption
{Evolution of the broadband 10 $\mu$m (N-N$_0$) and
20 $\mu$m (Q-Q$_0$) excesses as a function of time.
Left panels: 0.84--1.16 AU torus calculations;
right panels: 0.68--1.32 AU torus calculations.
Fragmentation produces large excesses relative to
a stellar photosphere at $t \sim 10^4$ to $10^5$ yr.
By $t \sim 10^6$ yr, the 10 $\mu$m excess is unobservable;
a small excess persists at 20 $\mu$m.}
\end{figure}
\clearpage

\begin{figure}
\plotone{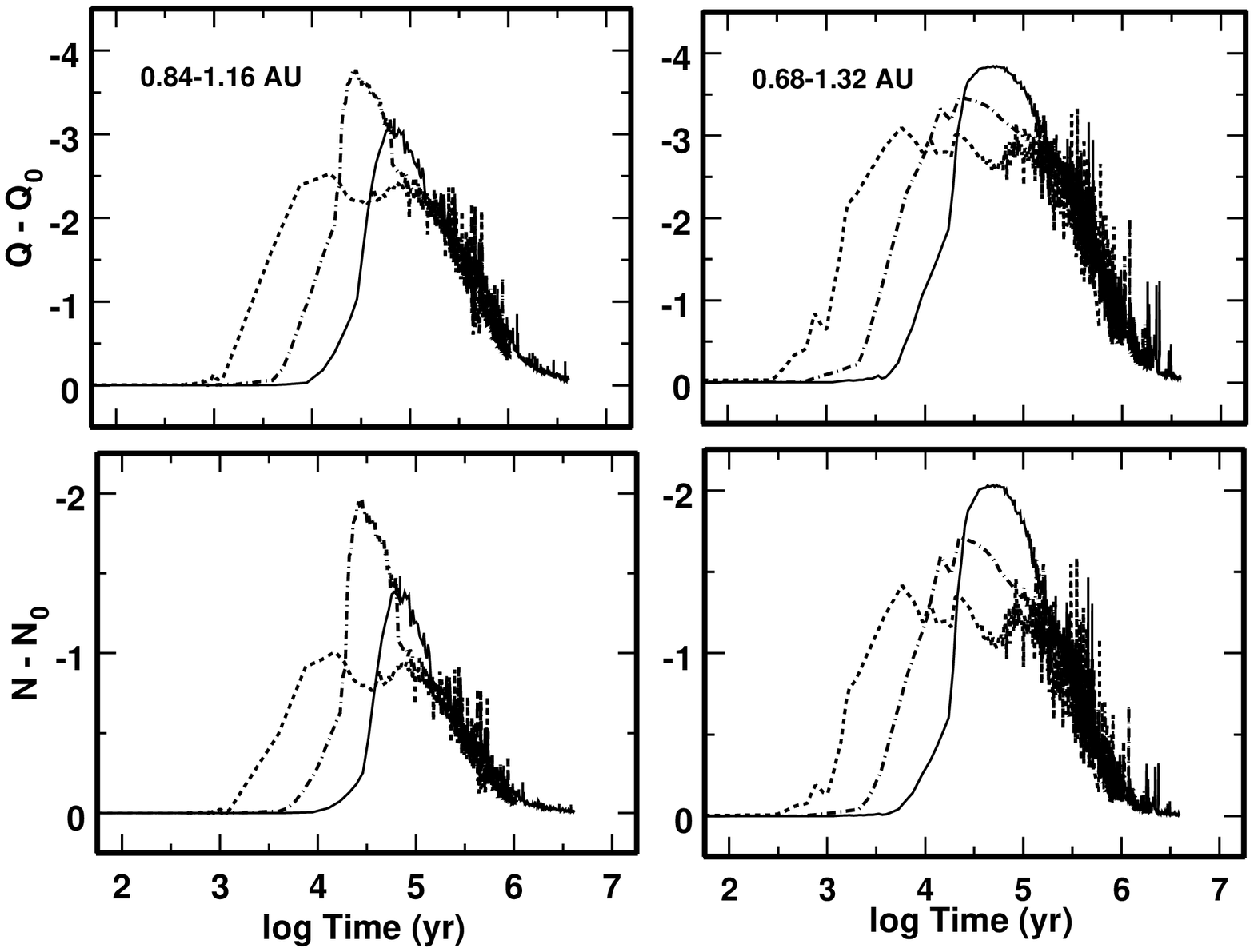} 
\caption
{Evolution of the broadband 10 $\mu$m (N-N$_0$) and
20 $\mu$m (Q-Q$_0$) excesses as a function of time.
Left panels: 0.84--1.16 AU torus calculations;
right panels: 0.68--1.32 AU torus calculations.
Fragmentation produces large excesses relative to
a stellar photosphere at $t \sim 10^4$ to $10^5$ yr.
By $t \sim 10^6$ yr, the 10 $\mu$m excess is nearly
unobservable; a small excess persists at 20 $\mu$m.}
\end{figure}


\begin{thebibliography}{99}

\vskip 4ex

\baselineskip=12pt
\parskip=0pt

\bibitem[Adachi et al. (1976)]{ada76} Adachi, I., Hayashi, C., \& Nakazawa, K.
1976, Progress of Theoretical Physics 56, 1756

\bibitem[Adams,~Lada, \& Shu (1987)]{ada87} Adams, F. C., Lada, C.J., 
\& Shu, F. H. 1987, ApJ, 308, 788 

\bibitem[Adams \& Watkins (1995)]{ada95} Adams, F. C., \& Watkins, R. 1995, ApJ, 451, 314

\bibitem[Agnor \& Ward (2002)]{agn02} Agnor, C. B., \& Ward, W. R. 2002,
ApJ, 567, 579

\bibitem[D'Alessio,~Calvet, \& Hartmann (2001)]{DA01}
D'Alessio, P., Calvet, N., \& Hartmann, L. 2001, ApJ, 553, 221

\bibitem[Bertout,~Basri, \& Bouvier (1988)]{ber88} Bertout, C., 
Basri, G., \& Bouvier, J. 1988, ApJ, 330, 350

\bibitem[van Boekel et al. (2003)]{vB03} van Boekel, R., Waters, L. B. F. M., 
Dominik, C., Bouwman, J., de Koter, A., Dullemond, C. P., \& Paresce, F. 2003,
A\&A, 400, L21

\bibitem[Bontemps et al. (2001)]{bon01} Bontemps, S. et al. 2001, A\&A, 372, 173

\bibitem[Bromley \& Kenyon (2004)]{bk04} Bromley, B., \& Kenyon, S. J.
2004, in preparation

\bibitem[Chambers (2001)]{cha01} Chambers, J. E. 2001, Icarus, 152, 205

\bibitem[Chiang \& Goldreich (1999)]{chi99}Chiang, E.I., \& Goldreich, P. 
1999, ApJ, 519, 279

\bibitem[Davis et al. (1985)]{dav85} Davis, D. R., Chapman, C. R.,
Weidenschilling, S. J., \& Greenberg, R. 1985, Icarus, 62, 30

\bibitem[Dominik \& Decin (2004)]{dom04} Dominik, C., \& Decin, G. 2004,
ApJ, in press (astro-ph/0308364)

\bibitem[Goldreich \& Ward (1973)]{gol73} Goldreich, P., \&
Ward, W. R. 1973,  ApJ, 183, 1051

\bibitem[Greaves \& Wyatt (2003)]{gre03} Greaves, J. S., \& Wyatt, M. C.
2003, MNRAS, in press

\bibitem[Habing et al. (2001)]{hab01} Habing, H. J., et al. 2001,
A\&A, 365, 545

\bibitem[Haghighipour \& Boss (2003)]{hag03} Haghighipour, N., \& Boss, A. P.
2003, ApJ, 583, 996
 
\bibitem[Haisch et al. (2001a)]{hai01a} Haisch, K. E., Jr., Lada, E. A., 
Pi\~na, R. K.; Telesco, C. M., \& Lada, C. J. 2001, AJ, 121, 1512

\bibitem[Haisch,~Lada, \& Lada (2001b)]{hai01b} Haisch, K., Lada, E. A.,
\& Lada, C. J. 2001, ApJ, 553, L153

\bibitem[Hartigan et al. (1990)]{har90} Hartigan, P., Hartmann, L., Kenyon, 
S. J., Strom, S. E., \& Skrutskie, M. F. 2001, ApJL, 354, L25
	
\bibitem[Hayashi (1981)]{hay81} Hayashi, C. 1981, Prog Theor Phys Suppl, 70, 35

\bibitem[Hogerheijde et al. (2003)]{hog03} Hogerheijde, M. R., Johnstone, D., 
Matsuyama, I., Jayawardhana, R., Muzerolle, J. 2003, ApJ, 593, L101

\bibitem[Hollenbach et al. (1994)]{hol94} Hollenbach, D. J., Johnstone, D., 
Lizano, S., \& Shu, F.  1994, ApJ, 428, 654

\bibitem[Jayawardhana et al. (1998)]{jay98} Jayawardhana, R. et al. 1998,
ApJ, 503, L79

\bibitem[Jayawardhana et al (2001)]{jay01} Jayawardhana, R., Wolk, S. J., 
Barrado y Navascu\~es, D., Telesco, C. M., \& Hearty, T. J. 2001, ApJ, 550, L197

\bibitem[Lagrange et al. (2000)]{lag00} Lagrange, A.-M., Backman, D.,
\& Artymowicz, P. 2000, in Protostars \& Planets IV, eds.  V. Mannings,
A. P. Boss, \& S. S. Russell, Tucson, Univ. of Arizona, p. 639

\bibitem[Kenyon \& Bromley (2001)]{kb01} Kenyon, S. J., \& Bromley, B. C.
2001, AJ, 121, 538

\bibitem[Kenyon \& Bromley (2002a)]{kb02a} Kenyon, S. J., \& Bromley, B. C., 
2002a, AJ, 123, 1757

\bibitem[Kenyon \& Bromley (2002b)]{kb02b} Kenyon, S. J., \& Bromley, B. C.
2002b, ApJ, 577, L35

\bibitem[Kenyon \& Bromley (2004)]{kb04} Kenyon, S. J., \& Bromley, B. C., 
2004, AJ, 127, No. 1

\bibitem[Kenyon \& Hartmann (1987)]{kh87} Kenyon, S. J., \&
Hartmann, L. 1987, ApJ, 323, 714

\bibitem[Kenyon \& Hartmann (1995)]{kh95} Kenyon, S. J., \& 
Hartmann, L.  1995, ApJS, 101, 117

\bibitem[Kenyon \& Luu (1998)]{kl98} Kenyon, S. J., \& Luu, J. X. 1998,
AJ, 115, 2136
 
\bibitem[Kenyon \& Luu (1999)]{kl99} Kenyon, S. J., \& Luu, J. X.
1999, AJ, 118, 1101

\bibitem[Kenyon et al. (1999)]{ken99} Kenyon, S. J., Wood, K.,
Whitney, B. A., \& Wolff, M. 1999, ApJ, 524, L119
 
\bibitem[Konigl \& Pudritz (2000)]{kon00} Konigl, A., \& Pudritz, R. 
2000, 2000, in Protostars \& Planets IV, eds.  V. Mannings,
A. P. Boss, \& S. S. Russell, Tucson, Univ. of Arizona, 759

\bibitem[Lissauer (1987)]{lis87} Lissauer, J. J. 1987, Icarus, 69, 249

\bibitem[Lissauer (1993)]{lis93} Lissauer, J. J. 1993, ARA\&A, 31, 129

\bibitem[Lynden-Bell \& Pringle (1974)]{lbp74} Lynden-Bell, D., \& 
Pringle, J. E. 1974, MNRAS, 168, 603

\bibitem[Najita,~Carr, \& Mathieu (2003)]{naj03} Najita, J., 
Carr, J. S., \& Mathieu, R. D. 2003, ApJ, 589, 931

\bibitem[Ohtsuki,~Stewart, \& Ida (2002)]{oht02} Ohtsuki, K., Stewart, 
G. R., \& Ida, S. 2002, Icarus, 155, 436

\bibitem[Ozernoy et al. (2000)]{oze00} Ozernoy, L. M., Gorkabyi, N. N.,
Mather, J. C., \& Taidakova, T. A. 2000, ApJ, 537, L147

\bibitem[Pollack et al. (1996)]{pol96} Pollack, J. B., Hubickyj, O.,
Bodenheimer, P., Lissauer, J. J., Podolak, M., \& Greenzweig, Y. 1996,
Icarus, 124, 62

\bibitem[Rafikov (2001)]{raf01} Rafikov, R. R. 2001, AJ, 122, 2713

\bibitem[Rafikov (2003)]{raf03} Rafikov, R. R. 2003, AJ, 126, 2529

\bibitem[Rucinski (1985)]{ruc85} Rucinski, S. M. 1985, AJ, 90, 2321
 
\bibitem[Safronov (1969)]{saf69} Safronov, V. S. 1969, {\it Evolution of 
the Protoplanetary Cloud and Formation of the Earth and Planets,}
Nauka, Moscow [Translation 1972, NASA TT F-677]

\bibitem[Schneider et al. (1999)]{sch99} Schneider, G., et al. 1999, ApJ, 513,
L127

\bibitem[Spangler et al. (2001)]{spa01} Spangler, C., Sargent, A. I., 
Silverstone, M. D., Becklin, E. E., \& Zuckerman, B. 2001, ApJ, 555, 932
 
\bibitem[Stassun et al. (2001)]{sta01} Stassun, K. G., M., Robert D., 
Vrba, F. J., Mazeh, T., \& Henden, A.  2001, AJ, 121, 1003

\bibitem[Supulver \& Lin (2000)]{sup00} Supulver, K. D., \& Lin, D. N. C. 2000, Icarus, 146, 525

\bibitem[Throop et al (2001)]{thr01} Throop, H. B., Bally, J., Esposito, L. W.,
\& McCaughrean, M. J. 2001, Science, 292, 1686

\bibitem[Weidenschilling (1977)]{wei77} Weidenschilling, S. J. 1977,
Ap Sp Sci, 51, 153

\bibitem[Weidenschilling \& Cuzzi (1993)]{wei93} Weidenschilling, S. J.,
\& Cuzzi, J. N., 1993, in Protostars and Planets III,
ed. E. H. Levy \& J. I. Lunine (Tucson: Univ of Arizona), p. 1031

\bibitem[Weidenschilling et al. (1997)]{wei97} Weidenschilling, S. J.,
Spaute, D., Davis, D. R., Marzari, F., \& Ohtsuki, K. 1997, Icarus, 
128, 429

\bibitem[Weinberger et al. (2003)]{wei03} Weinberger, A. J., Becklin, E. E.,
Zuckerman, B., Song, I. 2003, BAAS, 202, 3401

\bibitem[Weinberger et al. (1999)]{wei99} Weinberger, A. J., Becklin, E. E.,
Schneider, G., Smith, B. A., Lowrance, P. J., Silverstone, M. D.,
Zuckerman, B., Terrile, R. J. 1999, ApJ, 525, L53

\bibitem[Wetherill \& Stewart (1993)]{ws93} Wetherill, G. W., \& 
Stewart, G. R.  1993, Icarus, 106, 190

\bibitem[Williams \& Wetherill (1994)]{wil94} Williams, D. R., \&
Wetherill, G. W. 1994, Icarus, 107, 117

\bibitem[Wilner et al. (2002)]{wil02} Wilner, D. J., Holman, M. J.,
Kuchner, M. J., \& Ho, P. T. P. 2002, ApJ, 569, 115

\bibitem[Wood et al. (2002)]{woo02} Wood, K., Lada, C. J., Bjorkman, J. E.,
Kenyon, S. J., Whitney, B., Wolff, M. J. 2002, ApJ, 567, 1183

\bibitem[Wyatt,~Dent, \& Greaves (2003)]{wya03} Wyatt, M. C., Dent, W. R. F.,
\& Greaves, J. S. 2003, MNRAS, 342, 867

\bibitem[Youdin \& Shu (2003)]{you03} Youdin, A. N., \& Shu, F. H. 2002,
ApJ, ApJ, 580, 494

\end{thebibliography}
\end{document}